\documentstyle[epsf]{elsart}

\def\lQ{\Lambda_{\rm QCD}}
\newcommand{\nn}{\nonumber}
\newcommand{\be}{\begin{equation}}
\newcommand{\ee}{\end{equation}}
\newcommand{\bea}{\begin{eqnarray}}
\newcommand{\eea}{\end{eqnarray}}
\def\al{\alpha}
\def\als{\alpha_{\rm s}}
\def\siml{{\ \lower-1.2pt\vbox{\hbox{\rlap{$<$}\lower6pt\vbox{\hbox{$\sim$}}}}\ }} 

\begin{document}
\begin{frontmatter}
\begin{flushright}
\tt{CERN-TH/2000-205\\ UCSD/PTH 00-18}
\end{flushright}
\vskip 1truecm
\title{The Renormalization Group Improvement of the QCD Static Potentials}
\author {Antonio Pineda$^1$ and Joan Soto$^2$}
\address{$^1$ Theory Division, CERN, 1211 Geneva 23, Switzerland}
\address{$^2$ Department of Physics, University of California at San Diego,\\
     9500 Gilman Drive, La Jolla, CA 92093, USA}

\begin{abstract}
We 
resum the leading ultrasoft logs
of the singlet and octet
static QCD potentials within potential NRQCD. We then obtain the complete
three-loop renormalization group improvement of the singlet QCD static
potential. The discrepancies between the perturbative evaluation and the lattice results 
at short distances are slightly reduced.  
\end{abstract}

\vspace{1cm}
\end{frontmatter}

\newpage        

\pagenumbering{arabic}

Recently, there has been quite a lot of effort in identifying the large logs
that arise from the different scales involved 
in heavy quark systems near threshold \cite{current,NNLO,short,long,kp1,logs,kp2}. The proper renormalization group (RG)
 resummation of these logs is a non-trivial issue. It has recently been addressed within the so-called vNRQCD approach \cite{lmr}
for the $O(1/m^2)$ potentials \cite{ai1}, and for the $O(1/m)$ potential and the production current 
\cite{ai2}. 
Since the proper resummation of the logs may be important for the physics of
the top quark production near threshold,  
as well as 
for heavy quarkonium systems, and the existing calculations (see \cite{NNLO})
are done in a formalism closer to potential NRQCD (pNRQCD) \cite{Mont,long},
it would be desirable to know how to RG-improve within the latter formalism. 
In this note we 
provide the first step towards this goal by showing how the static potentials
of QCD can be RG-improved within that 
framework. Moreover, the RG-improved static potentials obtained in this paper
represent a new result by themselves. They provide the complete three-loop
RG evolution of the static potentials within an expansion
in $\als$ (in order to be so, besides our calculation, one needs to know the
static potentials at two loops \cite{Peter2,Schoeder2}). Our results may also
be relevant in order to understand the discrepancies, at relatively short distances, 
between the perturbative 
evaluation and the lattice results.

\medskip

Since we are only interested in the static potentials, we only need to consider
the static
limit of NRQCD \cite{NRQCD} and pNRQCD, i.e. we only need to work at leading order in
$1/m$. 
The matching between NRQCD and pNRQCD in
the static limit, in the situation where $\lQ \ll 1/r$ (the limit we will
consider in this paper),  
has been worked out in detail in ref.
\cite{long}.

The pNRQCD lagrangian at leading order in $1/m$ and next-to-leading order in the multipole expansion reads
\begin{eqnarray}
& & {\cal L}_{\rm pNRQCD} =
{\rm Tr} \,\Biggl\{ {\rm S}^\dagger \left( i\partial_0 - V_s  \right) {\rm S} 
+ {\rm O}^\dagger \left( iD_0 - V_o \right) {\rm O} \Biggr\}
\nonumber\\
\nonumber
& &\qquad + g V_A ( r) {\rm Tr} \left\{  {\rm O}^\dagger {\bf r} \cdot {\bf E} \,{\rm S}
+ {\rm S}^\dagger {\bf r} \cdot {\bf E} \,{\rm O} \right\} 
+ g {V_B (r) \over 2} {\rm Tr} \left\{  {\rm O}^\dagger {\bf r} \cdot {\bf E} \, {\rm O} 
+ {\rm O}^\dagger {\rm O} {\bf r} \cdot {\bf E}  \right\}  
\\
& &\qquad- {1\over 4} F_{\mu \nu}^{a} F^{\mu \nu \, a}+O(r^2)\,.
\label{pnrqcd0}
\end{eqnarray}
All the gauge fields in Eq. (\ref {pnrqcd0}) are evaluated 
in ${\bf R}$ and $t$, in particular $F^{\mu \nu \, a} \equiv F^{\mu \nu \, a}
({\bf R},t)$ and $iD_0 {\rm O} \equiv i \partial_0 {\rm O} - g [A_0({\bf
  R},t),{\rm O}]$ and
$$
V_s=- C_f{\al_{V_s} \over r}
$$
$$
V_o=\left({C_A \over 2} -C_f\right){\al_{V_o} \over r}
.
$$
The potentials
$V_{i}$, $i=s,o,A,B$ are to be regarded as matching coefficients, which depend
on the scale $\nu$ separating soft gluons from ultrasoft ones. In the static
limit, 
we understand by soft energies the ones of $O(1/r)$ and by ultrasoft energies
the ones of $O(\als/r)$. 
Notice that the hard scale, $m$, plays no role in the static limit.
The only assumption 
made so far 
concerning the size of $r$ is that $1/r \gg \lQ$. 
However, in order to use perturbative RG techniques, we shall also assume from now on that we are working at scales $\nu$ 
such that
$\als(\nu) \ll 1$.

Formally, we can write Eq. (\ref{pnrqcd0}) as an expansion in $r$ in the following way:
\be
{\cal L}_{\rm pNRQCD} =\sum_{n=-1}^{\infty}r^n\lambda_n^BO_n^B
, 
\ee
where the above fields and parameters should be understood as bare and the renormalization group equations of the renormalized matching
coefficients read 
\be
\nu {d \over d \nu}\lambda=B_{\lambda}(\lambda)
.
\ee
If $\lambda_{-1}=0$,
there are no
relevant operators (superrenormalizable terms) in the Lagrangian
and the RG equations have a triangular structure (the standard
structure one can see, for instance, in HQET \cite{hqet}; see Ref. \cite{Neubert} for a
review): 
\be
\nu {d \over d \nu}\lambda_0=B_0(\lambda_0)
\ee
\be
\nu {d \over d \nu}\lambda_1=B_1(\lambda_0)\lambda_1
\ee
\be
\nu {d \over d \nu}\lambda_2=B_{2,(2,1)}(\lambda_0)\lambda_2+B_{2,(1,2)}(\lambda_0)\lambda_1^2
\ee
\be
....\,,
\ee
where the different B's can be power-expanded in $\lambda_0$ ($\lambda_0$
corresponds to the marginal operators (renormalizable interactions)). 
If $\lambda_{-1}\not= 0$,
however, there are
relevant operators (superrenormalizable terms) in the Lagrangian
and the RG equations lose the triangular structure. Still,
if $\lambda_{-1} \ll 1$, a perturbative calculation of the renormalization
group equations can be achieved as a double expansion in $\lambda_{-1}$ and
$\lambda_0$. The RG equations now have the following structure: 
\bea
&&\nu {d \over d \nu}\lambda_{-1}
\\
&&
\nn
\quad
=B_{-1}(\lambda_0)\lambda_{-1}+B_{(-1,2)}(\lambda_0)\lambda_{-1}^2\lambda_1+B_{(-1,3)}^{(a)}(\lambda_0)\lambda_{-1}^3\lambda_1^2+B_{(-1,3)}^{(b)}(\lambda_0)\lambda_{-1}^3\lambda_2+O(\lambda_{-1}^4)
\eea
\be
\nu {d \over d \nu}\lambda_0=B_0(\lambda_0)+B_{0,1}(\lambda_0)\lambda_{-1}\lambda_1+O(\lambda_{-1}^2)
\ee
\be
\nu {d \over d \nu}\lambda_1=B_1(\lambda_0)\lambda_1+B_{1,1}(\lambda_0)\lambda_{-1}\lambda_1^2+O(\lambda_{-1}^2)
\ee
and so on (for a similar discussion in the context of scalar $\lambda \phi^n$-like
theories 
see \cite{AtanceCortes}).

At short distances, the static limit of pNRQCD lives in the second situation. Specifically, we
have $\lambda_{-1}=\{\al_{V_s},\al_{V_o}\}$, that fulfils $\lambda_{-1} \ll 1$;
$\lambda_0=\al_s$  and $\lambda_{1}=\{V_A,V_B\}$. Therefore, 
we can calculate the anomalous dimensions order by
order in $\als$. In addition, we also have an expansion in $\lambda_{-1}$ that
corresponds to working order by order in the multipole expansion.

The specific form of the pNRQCD lagrangian
severely constrains the above general structure.
From ref. \cite{long}\footnote{The result of the last two equations corrects the previous
  evaluation of Ref. \cite{long} (we thank A. Vairo for confirming these results). In order to cross-check this result we have redone
  the computation both in the Coulomb gauge and in the
  background field gauge.}
we obtain at leading,
non-vanishing, order

\bea
\nu {d\over d\nu}\al_{V_{s}}
&=& 
{2 \over 3}{\alpha_{s}\over
  \pi}V_A^2\left( \left({C_A \over 2} -C_f\right)\al_{V_o}+C_f\al_{V_s}\right)^3
+O(\lambda_{-1}^4\lambda_0,\lambda_0^2\lambda_{-1}^3) 
\nn \\
\label{RGeq}
\nu {d\over d\nu} \al_{V_{o}}&=& {2 \over 3} {\alpha_{s}\over \pi}
V_A^2\left(\left({C_A \over 2} -C_f\right)\al_{V_o}+C_f\al_{V_s}\right)^3 +O(\lambda_{-1}^4\lambda_0,\lambda_0^2\lambda_{-1}^3) 
\nn\\
\nu {d\over d\nu}\alpha_{s}&=&\alpha_{s}\beta(\alpha_{s})
\nn\\
\nu {d\over d\nu} V_{A}&=& 0
+O(\lambda_{-1}\lambda_0,\lambda_0^2) 
\\
\nu {d\over d\nu} V_{B}&=& 0
+O(\lambda_{-1}\lambda_0,\lambda_0^2)  \nn
\,.
\eea

It is the aim of this work to solve Eq. (\ref{RGeq}), and hence to provide the leading log (LL) ultrasoft RG improvement of the 
pNRQCD lagrangian in the static limit.

The last two equations in Eq. (\ref{RGeq}) are decoupled from the rest and are
equal to zero
at the order we are working. Therefore, we immediately obtain\footnote{Throughout the paper, we use
$\alpha_s (r)$ as a shorthand for $\alpha_s (\nu =1/r)$ and similarly
for any other running object at the scale $\nu =1/r$.}
\be
 V_{A}(\nu) = V_{A}({r}) \qquad {\rm and} \qquad V_{B}(\nu) = V_{B}({ r})
.
\ee

The first two equations of Eq. (\ref{RGeq}) are equivalent to
\bea
\nu {d\over d\nu} \left(\left({C_A \over 2} -C_f\right)\al_{V_o}+C_f\al_{V_s}\right)&=& \gamma_{os}{\alpha_{s}\over \pi}V_A^2\left(\left({C_A \over 2} -C_f\right)\al_{V_o}+C_f\al_{V_s}\right)^3 \nn \\
\nu {d\over d\nu} \left(\al_{V_o}-\al_{V_s}\right)&=&0
,
\eea
where
\be
\gamma_{os}= {C_A\over 3}
.
\ee
The second equation tells us that the combination $\al_{V_o}-\al_{V_s}$ is 
RG-invariant at this order. The first equation can now be easily solved (note
that in order to solve these equations with the demanded accuracy it was
necessary to know the RG solution of $V_A$, which happened to be
trivial). We obtain (where $\beta_0=11 C_A / 3 - 4 T_F n_f / 3$, $T_F=1/2$ and
$n_f $ is the number of light quarks): 
\bea
\al_{V_{s}}(\nu )&=&\al_{V_{s}}(r)-{2\over 3 \gamma_{os}} \left(\left({C_A \over 2} -C_f\right)\al_{V_o}(r)+C_f\al_{V_s}(r) \right) \nn \\
& & \times \left(
1- {1\over \sqrt{1-{4\gamma_{os} \over  \beta_0} \left(\left({C_A \over 2} -C_f\right)\al_{V_o}(r)+C_f\al_{V_s} ({ r})\right)^2  
 V_{A}^2({ r})\log\left(
\alpha_{s}({r})\over \alpha_{s}(\nu) \right) }}
\right) \nn\\
\label{sol}
\al_{V_{o}}(\nu )&=&\al_{V_{o}}(r)-{2\over 3 \gamma_{os}} \left(\left({C_A \over 2} -C_f\right)\al_{V_o}(r)+C_f\al_{V_s}(r) \right)  \\
\nn
& & \times \left(
1- {1\over \sqrt{1-{4\gamma_{os} \over  \beta_0} \left(\left({C_A \over 2} -C_f\right)\al_{V_o}(r)+C_f\al_{V_s} ({ r})\right)^2  
 V_{A}^2({ r})\log\left(
\alpha_{s}({r})\over \alpha_{s}(\nu) \right) }}
\right) 
, 
\eea
which completes our results. These are exact up to
$O\left(\lambda_{-1}^3(r)\left(\lambda_{-1}^2(r)\left(\lambda_0(r)\log{r\nu}\right)^n
\right)^m,\right.\\
\left.
\lambda_{-1}(r)\lambda_0(r)\left(\lambda_{-1}^2(r)\left(\lambda_0(r)\log{r\nu}\right)^n\right)^m\right) 
$ where $n=0,1,2...$ and $m=1,2...\,$.    
At this point we would like to stress the
simplicity of the calculation, which follows to a large extent from the formalism
used.
\begin{center}
\begin{figure}
\epsfbox{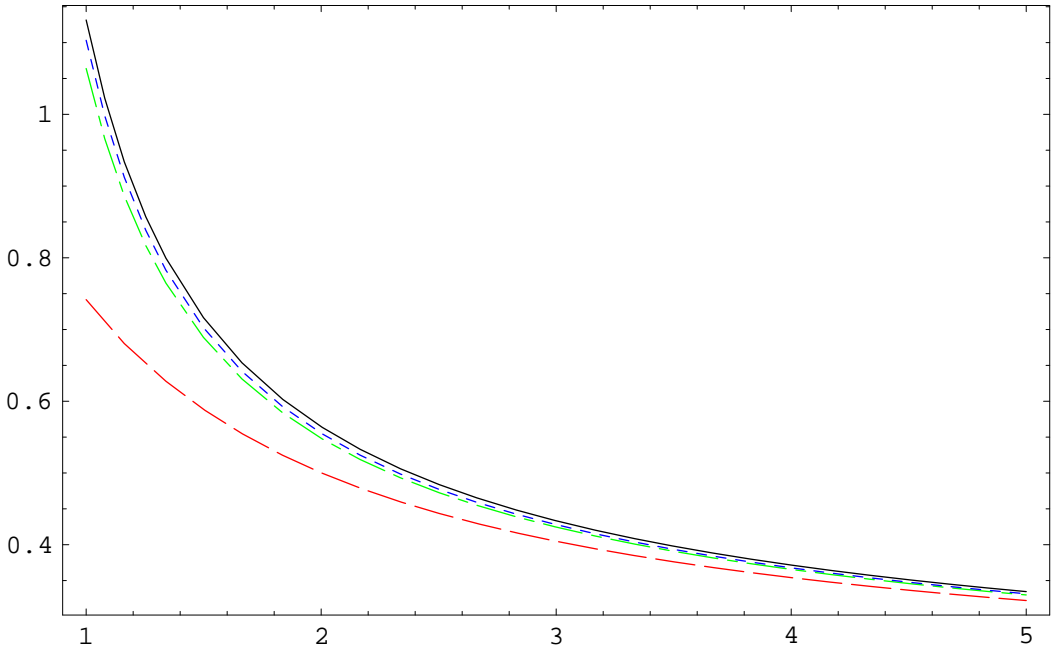}
\vspace{0.2cm}
\caption{We plot $\al_{V_s}$ versus $r$ (GeV) for a range of energies  
  relevant to a comparison with lattice simulations and also to the study of
  the $\Upsilon(1S)$ system ($n_f=4$, $\lQ^{{\rm three\,loops}}=0.28$ GeV). The solid line corresponds to
  $\al^{{\rm (2\,loops)}}_{V_s}$, the dotted line to $\al_{V_s}^{{\rm(2\,
  loops)}}$ plus  
  the single leading-log correction computed in Ref. \cite{short}, the
  dot-dashed line to Eq. (\ref{solals}) and the dashed line to
  Eq. (\ref{sol}). We have fixed $\nu=\als(r)/r$.} 
\end{figure}
\end{center}

In the above results the different matching coefficients can be considered to
be independent (though assuming $\lambda_{-1}$, $\lambda_0 \ll 1$ in order for 
our perturbative evaluation to make sense). If we want to perform a strict expansion
in $\als$, one has to use the perturbative relation between $\lambda_{-1}(r)$, $\lambda_{1}(r)$
and $\als(r)$: $\lambda_{-1}(r)=A_1\als(r)+A_2\als^2(r)+...$ and
$\lambda_{1}(r)=1+B_1\als(r)+...$.  
Since $\al_{V_{s}}$ and $\al_{V_o}$ are known at two-loop accuracy \cite{Peter2,Schoeder2}, Eq. (\ref{sol}) 
is known with the following accuracy

\bea
\label{solals}
\al_{V_{s}}(\nu )&=&\al_{V_{s}}^{(2\, {\rm loops})}({ r})+{C_A^3\over
  6\beta_0}\als^3(r) \log\left(
\alpha_{s}({r})\over \alpha_{s}(\nu) \right) +O(\als(r)^{4+n}\log^n{r \nu})\\
\nn
\al_{V_{o}}(\nu )&=&\al_{V_{o}}^{(2\, {\rm loops})}({ r})+{C_A^3\over
  6\beta_0}\als^3(r) \log\left(
\alpha_{s}({r})\over \alpha_{s}(\nu) \right) +O(\als(r)^{4+n}\log^n{r \nu})
,
\eea
where $n=0,1,2,...\,$.

The above equation represents the complete three-loop RG
improvement of the static potentials. 
When the running $\alpha_s (r)$ is substituted by its three-loop expression
above and expanded, all the $\log r \nu $ in the three-loop static potentials
are obtained, including the infrared logs of \cite{adm,short}. More explicitly, if we write 
$\lambda_{-1}=\lambda_{-1}(r,\nu,\als(\nu))$, Eq. (\ref{solals}) provides the
correct $\als^{1+n}(\nu)\log^n{r \nu}$, $\als^{2+n}(\nu)\log^n{r \nu}$,
$\als^{3+n}(\nu)\log^n{r \nu}$ terms at any power of $n$. Note that such
accuracy is not achieved by only replacing $\alpha_s (r)$ by its three-loop running expression
in the two-loop calculation of the potential, as it is sometimes done in the
literature \cite{Peter1} (see also \cite{russos}).    

The Lagrangian (\ref{pnrqcd0}) is suitable for the study of static systems. In
   particular, we may be interested in computing the energy of two
   infinitely heavy (static) sources separated at a distance $r$ in a singlet configuration. This is,
   indeed, the object computed in lattice simulations and usually referred as
   the static potential. Within our formalism, this energy corresponds to the pole of the
   singlet-singlet correlator. In the energy, the scale dependence $\nu$ of $V_s$ gets cancelled by ultrasoft contributions of $O(
   \als(r)/r)$. Therefore, by lowering the scale $\nu$ in $V_s$ to values of $O(
   \als(r)/r)$, all the large logs arising from the ratio between the soft and
   ultrasoft scale are resummed and the energy becomes, at the accuracy we are
   working, equal to $V_s$ with  $\nu=\als(r)/r$ (up to finite
   terms). 
Now we would like to give some numerical estimates of our results (we
fix $\nu=\als(r)/r$ in what follows).  
In Fig. 1, we plot $\al_{V_s}$ versus $r$ (GeV), within different
  approximations.
If we compare the last
  term in Eq. (\ref{solals}) with the single leading-log result obtained
  in \cite{short}, we can see a sizeable correction of the same order of
  magnitude. If for definiteness we choose $r^{-1}=2\, {\rm GeV}$,
  Eq. (\ref{solals}) gives a correction of around 3\%, whereas the single 
  leading-log term gives a correction of around 1\%. This
  shows the necessity to RG-improve the single leading logs. In any case, as we can see from Fig. 1 and the above results, the
  correction is rather tiny whith respect to the total value of
  $\al_{V_S}$. 
Equation (\ref{sol}), however, produces a much more sizeable correction, as we can see
  in Fig. 1. If we again take $r^{-1}=2\, {\rm GeV}$ the correction is of
  around  11\%.

In all the cases the new results lower the value of $\al_{V_s}$, 
  reducing the existing discrepancies between the perturbative results and the
  lattice simulations \cite{Kuti}. This is specially so when using
  Eq. (\ref{sol}).  
Although 
the good behaviour of $\al_{V_s}$ using Eq. (\ref{sol}) is certainly appealing, we should 
bear in mind that around the $1\, {\rm GeV}$ region $\lambda_{-1}(r)\sim 1$ 
and hence perturbation theory becomes 
unreliable. This is so specially in the perturbative relation between
$\al_{V_s}(r)$ and $\als(r)$ and it is usually associated to renormalon
effects \cite{pertren}. This point should be better understood before a
detailed study about the impact of the RG improvement can be done.

Since for $\al_{V_o}(r)$ only the
  one-loop expression is available, we refrain from performing a
  similar study for it (note that in the numerical analysis above we have used
  the one-loop expression of $\al_{V_o}(r)$ in Eq. (\ref{sol})). 

One could also consider non-perturbative effects in the static case when $1/r \gg \lQ$ (see \cite{long}
 for a more general discussion). Here we only mention that when $\alpha_s
 (r)/r \gg \lQ$ the leading non-perturbative contribution is proportional to $
 r^3 \langle\alpha_s FF\rangle/\alpha_s (r) $ \cite{NP}, whereas for $\alpha_s (r)/r \sim
 \lQ$ one has a non-perturbative contribution proportional to $r^2$ times a
 non-local correlator of two chromoelectric fields 
 with multiple potential insertions of $O(\als(r)/r)$.  

Our results are also relevant to  
actual physical systems composed of a quark and an antiquark with
very large but finite mass. However, in order to achieve, for instance, an $O(m\als^{4+n}\log^n \als )$ accuracy in the binding 
energy of those systems, the RG improvement of the $1/m$ and $1/m^2$ potentials is also necessary. Since local 
field redefinitions allow to reshuffle contributions from a given order in $1/m$ to another \cite{1om}, a meaningful
 outcome can only be obtained when all the contributions from the various potentials are taken into account. Hence, we 
restrict ourselves here to present a few numerical estimates on the impact of
the RG-improved static potentials for some typical scales one can find in
the $\Upsilon(1S)$
and $t$-$\bar t$ systems. In particular, for the $\Upsilon(1S)$, for which
typical soft scales are of $O(2\,{\rm GeV})$, the discussion of Fig. 1 applies. For
$t$-$\bar t$ systems, in order to give an estimate, we take $r^{-1}=20\,{\rm GeV}$
($n_f=5$, $\lQ^{{\rm three\,loops}}=0.2$). We obtain that the corrections are
of order 0.4\%, 0.5\%, 1\%, if we consider the single leading-log
result, Eq. (\ref{solals}) and Eq. (\ref{sol}), respectively.

Before closing, let us mention that 
in the vNRQCD approach
it appears to be crucial that ultrasoft and soft gluons run from a scale $m v^2$ and $m v$, respectively, to the scale $m$.
In the static system, the scale $m$ does not exist, and hence it is not clear to us how one should proceed in the above-mentioned approach.
Notice also that the difficulties pointed out in \cite{aij} concerning a na\"\i ve RG improvement in two stages do not apply to the 
static case, 
since here only one stage is involved, namely the RG improvement between NRQCD and pNRQCD.

\bigskip

{\bf Acknowledgements} 

We thank M. Beneke, J. Kuti, A. Manohar, I. Stewart and A. Vairo for
stimulating discussions. We also thank G. Bali for sharing his lattice data
with us. 
A.P. acknowledges the TMR contract No. ERBFMBICT983405 and 
J.S. the financial support from the CICYT (Spain), contract AEN98-0431, the CIRIT (Catalonia), contract 
1998SGR 00026, and the grant BGP-08 (Catalonia). J. S. thanks A. Manohar and the HEP group at UCSD for their warm hospitality
 while this work was carried out.



\begin{thebibliography}{99}


\bibitem{current} M. Beneke, A. Signer and V.A. Smirnov,
      Phys. Rev. Lett. {\bf 80} (1998) 2535;  A. Czarnecki and K. Melnikov,
      Phys. Rev. Lett. {\bf 80} (1998) 2531.

\bibitem{NNLO} A.H. Hoang, M. Beneke, K. Melnikov, T. Nagano, A. Ota,
  A.A. Penin, A.A. Pivovarov, A. Signer, V.A. Smirnov, Y. Sumino, T. Teubner,
  O. Yakovlev and A. Yelkhovsky, Eur. Phys. J. direct {\bf C 3} (2000) 1.

\bibitem{short} N. Brambilla, A. Pineda, J. Soto and A. Vairo,
     Phys. Rev. {\bf D 60} (1999) 091502.

\bibitem{long} N. Brambilla, A. Pineda, J. Soto and A. Vairo,
     Nucl. Phys. {\bf B 566} (2000) 275. 

\bibitem{kp1} B. A. Kniehl and A. A. Penin,
      Nucl. Phys. {\bf B 563} (1999) 200.

\bibitem{logs}  N. Brambilla, A. Pineda, J. Soto and A. Vairo,
     Phys. Lett. {\bf B 470} (1999) 215.

\bibitem{kp2} B. A. Kniehl and A. A. Penin, hep-ph/9911414. 

\bibitem{lmr} M. E. Luke, A. V. Manohar and I. Z. Rothstein,
      Phys. Rev. {\bf D 61} (2000) 074025.

\bibitem{ai1}  A. V. Manohar and I. W. Stewart, hep-ph/9912226; hep-ph/0003032.

\bibitem{ai2} A. V. Manohar and I. W. Stewart, hep-ph/0003107.

\bibitem{Mont} A. Pineda and J. Soto,
      Nucl. Phys. Proc. Suppl. {\bf 64} (1998) 428.

\bibitem{Peter2} Y. Schr\"oder,
      Phys. Lett. {\bf B 447} (1999) 321;  M. Peter,
      Nucl. Phys. {\bf B 501} (1997) 471.

\bibitem{Schoeder2} Y. Schr\"oder, private communication.

\bibitem{NRQCD} W. E. Caswell and G. P. Lepage, Phys. Lett. {\bf B 167} (1986)
  437; G.T. Bodwin, E. Braaten and G.P. Lepage, Phys. Rev. {\bf D 51}
  (1995) 1125, Erratum {\it ibid.} {\bf D 55} (1997) 5853.

\bibitem{hqet}  S. Balk, J.G. Korner and D. Pirjol, Nucl. Phys. {\bf B 428} (1994) 499; 
C. Bauer and A. V. Manohar, Phys. Rev. {\bf D 57} (1998) 337; 
C. Balzereit, Phys. Rev. {\bf D 59} (1999) 034006.

\bibitem{Neubert} M. Neubert, Phys. Rep. {\bf 245} (1994) 259. 

\bibitem{AtanceCortes} M. Atance and J. L. Cortes, Phys. Rev. {\bf D 56} (1997)
  3611. 
 
\bibitem{adm} T. Appelquist, M. Dine and I. J. Muzinich, Phys. Rev. {\bf D 17} (1978) 2074.

\bibitem{Peter1} M. Peter, Phys. Rev. Lett. {\bf 78} (1997) 602.

\bibitem{russos} V. V. Kiselev, A. E. Kovalsky and A. I. Onishchenko, hep-ph/0005020.

\bibitem{Kuti}  
G.~S.~Bali and K.~Schilling,
Phys.\ Rev.\  {\bf D47}  (1993) 661; 
G.~S.~Bali and K.~Schilling,
Int.\ J.\ Mod.\ Phys.\  {\bf C4} (1993) 1167; 
G.~S.~Bali, K.~Schilling and A.~Wachter,
Phys.\ Rev.\  {\bf D56} (1997) 2566;
K. J. Juge, J. Kuti and C.J. Morningstar,  hep-lat/9911007.

\bibitem{pertren} M. Jezabek, J.H. Kuhn, M. Peter, Y. Sumino and T. Teubner,
  Phys. Rev. {\bf D 58} (1998), 014006;  M. Jezabek, M. Peter and Y. Sumino,
  Phys. Lett. {\bf B 428} (1998) 352. 

\bibitem{1om} N. Brambilla, A.Pineda, J. Soto and A. Vairo, hep-ph/0002250.

\bibitem{NP} C.A. Flory, Phys. Lett. {\bf B 113} (1982) 263; R.A. Bertlmann
  and J.S. Bell, Nucl. Phys. {\bf B 227} (1983) 435.

\bibitem{aij}  A. V. Manohar, J. Soto and I. W. Stewart, hep-ph/0006096.

\end{thebibliography}
\end{document}